 \documentclass[final,3p,times,twocolumn]{elsarticle}
\usepackage{graphics}
\usepackage{graphicx}
\usepackage{epsfig}
\usepackage{epstopdf}
\usepackage{amssymb}
\journal{Solid State Communications}

\begin{document}

\begin{frontmatter}

%% Title, authors and addresses

%% use the tnoteref command within \title for footnotes;
%% use the tnotetext command for the associated footnote;
%% use the fnref command within \author or \address for footnotes;
%% use the fntext command for the associated footnote;
%% use the corref command within \author for corresponding author footnotes;
%% use the cortext command for the associated footnote;
%% use the ead command for the email address,
%% and the form \ead[url] for the home page:
%%
%% \title{Title\tnoteref{label1}}
%% \tnotetext[label1]{}
%% \author{Name\corref{cor1}\fnref{label2}}
%% \ead{email address}
%% \ead[url]{home page}
%% \fntext[label2]{}
%% \cortext[cor1]{}
%% \address{Address\fnref{label3}}
%% \fntext[label3]{}

\title{Two-dimensional cavity polaritons under the influence of the perpendicular strong magnetic and electric fields. The gyrotropy effects}

%% use optional labels to link authors explicitly to addresses:
%% \author[label1,label2]{<author name>}
%% \address[label1]{<address>}
%% \address[label2]{<address>}

\author[rvt]{S.A.Moskalenko}
\author[rvt]{I.V.Podlesny}
\author[rvt]{E.V.Dumanov\corref{cor1}}
\ead{dum@phys.asm.md}
\cortext[cor1]{Corresponding author}
\author[foc]{M.A.Liberman}
\address[rvt]{Institute of Applied Physics of the Academy of Sciences of Moldova, Academic
Str. 5, Chisinau, MD2028, Republic of Moldova}
\address[foc]{Nordita, KTH Royal Institute of Technology and Stockholm University, Roslagstullsbacken 23, 10691 Stockholm, Sweden}

\begin{abstract}
The properties of the two-dimensional cavity polaritons subjected to the action of a strong perpendicular magnetic and electric fields, giving rise to the Landau quantization (LQ) of the 2D electrons and holes accompanied by the Rashba spin-orbit coupling, by the Zeeman splitting and by the nonparabolicity of the heavy-hole dispersion law are investigated. We use the method proposed by Rashba [1] and the  obtained results are based on the exact solutions for the eigenfunctions and for the eigenvalues of the Pauli-type Hamilonians with third order chirality terms and nonparabolic dispersion law for heavy-holes and with the first order chirality terms for electrons.
The selection rules of the band-to-band optical quantum transitions as well as of the quantum transitions from the ground state of the crystal to the magnetoexciton states depend essentially on the numbers $n_{e}$ and $n_{h}$ of the LQ levels of the (e-h) pair forming the magnetoexciton. It is shown that the Rabi frequency $\Omega_{R}$ of the polariton branches and the magnetoexciton oscillator strength $f_{osc}$ increase with the magnetic field strength $B$ as $\Omega_{R}\sim \sqrt{B}$, and $f_{osc}\sim B$. The optical gyrotropy effects may be revealed changing the sign of the photon circular polarization at a given sign of the wave vector longitudinal projection $k_{z}$ or eqivalently changing the sign of $k_{z}$ at the same selected circular polarization.
\end{abstract}
\end{frontmatter}

%%
%% Start line numbering here if you want
%%
% \linenumbers

%% main text
\section{Introduction}
% Put \label in argument of \section for cross-referencing
%\section{\label{}}
%\subsection{}
%\subsubsection{}
The long standing efforts of many groups of investigators were crowned by the experimental realization of the Bose-Einstein condensation (BEC) phenomenon in a system of two-dimensional (2D) cavity polaritons. The results of the theoretical and experimental studies in this field can be found in  Refs. [2-8]. Due to a very small effective mass of the cavity polaritons on the lower polariton branch, that is by four orders of magnitude smaller than that of the free electron mass, the BEC of cavity polaritons in GaN crystals was realized at room temperatures. The polariton laser in the form of a diode with electron injection was constituted on this basis [9]. The influence of the external magnetic field on the cavity polaritons and on their BEC was investigated for the first time in [10-14]. But these studies were concerned the relatively small strength of the magnetic field, so that the electron structure of the Wannier-Mott excitons remains unchanged, causing only the diamagnetic and Zeeman splitting effects. Another limit of magnetic field strength was studied in [15-17]. The structure of the magnetoexcitons in the GaN-type quantum wells (QWs) in a wide range of the perpendicular magnetic field up to 60T was investigated. In Section 2 we  consider the 2D electron-hole (e-h) system in GaAs-type QW structure with one heavy-hole valence subband using the rising and lowering operators in the space of the single-particle Landau levels. We use the exact solutions of the Landau quantization  taking into account the Rashba spin-orbit coupling (RSOC), Zeeman splitting (ZS) and non-parabolicity of the heavy-hole dispersion law  obtained in [18-21], and the corresponding electron and hole wave functions determined using the Rashba method [1].
 The Hamiltonian of the Coulomb electron-electron and of the electron-radiation interactions in the second quantization representation are obtained in Section 3 using these wave functions. On this basis the magnetoexciton wave functions and energy spectrum are determined. The Hamiltonian of the free magnetoexcitons and photons as well as of their interaction are introduced. It makes possible to turn from the electron-hole representation to the magnetoexciton-photon representation, to introduce in Section 4 the notion of polaritons following the Hopfield coherent superposition state [22]. We conclude in Section 5.

\section{Landau quantization, Rashba spin-orbit coupling and Zeeman splitting of the 2D electron and hole gases.}
The Hamiltonians describing the Landau quantization, Rashba spin-orbit coupling and Zeeman splitting effects with participation of the 2D electrons and heavy holes were derived in Refs.[18-21]. In dimensionless form they are:
\begin{eqnarray}
&H_{e}=\frac{H_{e}}{\hbar {{\omega }_{ce}}}=\{(a^{\dagger }a+{1}/{2}\;)\hat{I}+i\alpha \sqrt{2}\left| \begin{array}{cc}
   0 & a  \\
   -a^{\dagger } & 0  \\
\end{array} \right|+ &\nonumber \\
&+Z_{e}\hat{\sigma }_{z}\},&\nonumber \\
&H_{h}=\frac{H_{h}}{\hbar {\omega }_{ch}}=\{[(a^{\dagger }a+{1}/{2}\;)+\delta (a^{\dagger }a+{1}/{2}\;)^{2}]\hat{I}+& \nonumber\\
&+i\beta 2\sqrt{2}\left| \begin{array}{cc}
   0 & (a^{\dagger })^{3}  \\
   -a^{3} & 0  \\
\end{array} \right|+Z_{h}\hat{\sigma }_{z}\}&
\end{eqnarray}
Here the Bose operators $a^{\dagger },a$ generating the Fock oscillator-type states $\left| m \right\rangle $ were introduced, and we use the following notations
\begin{eqnarray}
&Z_{i}=\frac{{g_{i}}{\mu }_{B}B}{2\hbar {\omega }_{ci}}=\frac{{g_{i}}{m_{i}}}{4m_{0}},{\omega }_{ci}=\frac{eB}{{m_{i}}c},i=e,h, &\\
&\hat{I}=\left| \begin{array}{cc}
   1 & 0  \\
   0 & 1  \\
\end{array} \right|,{{{\hat{\sigma }}}_{z}}=\left| \begin{array}{cc}
   1 & 0  \\
   0 & -1  \\
\end{array} \right|, &\nonumber \\
&{{\mu }_{B}}=\frac{e\hbar }{2{{m}_{0}}c},e=|e|>0 &\nonumber
\end{eqnarray}
where ${\omega }_{ci}$ are the cyclotron frequencies, $Z_{i}$ are the Zeeman parameters proportional to the g-factors ${{g}_{i}}$ and to the effective masses $m_{i}$ of the electrons and holes, and $m_{0}$ is the bare electron mass. The parameters $\alpha $ and $\beta $ determine the role of the RSOC expressed through the first order chirality terms as was introduced for conduction electrons in Ref.[1] and by third order chirality terms as was established in Ref[23]. In Refs.[18-21] these parameters were written as functions of dimensionless values of the external  strong magnetic field perpendicular to the layer and electric field as following: $B=\xi $Tesla, ${{E}_{z}}=\eta $kV/cm, $\alpha =8\cdot {{10}^{-3}}\eta /\sqrt{\xi }$, $\beta=1.06\cdot {{10}^{-2}}\eta \sqrt{\xi }$. Similarly the parameter $\delta $ determining the nonparabolicity of the hh dispersion law expressed by the quadratic term in the Hamiltonian $H_{h}$ is written in the form: $\delta =10^{-4}C\xi \eta $ with unknown coefficient   C. Notice, that it was introduced in [18-21] to avoid the collapse of the semiconductor energy gap $E_{g}^{0}$ in the presence of the third order chirality terms, which  without quadratic term could lead to the unlimited penetration of the heavy hole energy levels inside the energy gap $E_{g}^{0}$. The exact solutions for the eigenvalues and eigenfunctions of both Hamiltonians (1) were obtained in Refs[18-21] using the Rashba method [1]. In the case of conduction electron they coincide with the Rashba solutions being supplemented by the ZS effects as follows:
\begin{eqnarray}
&\frac{E_{e,m}^{\pm }}{\hbar {{\omega }_{ce}}}=\varepsilon _{e,m}^{\pm }=(m+1)\pm \sqrt{{{\left( \frac{1}{2}-{{Z}_{e}} \right)}^{2}}+2|\alpha {{|}^{2}}(m+1)},& \nonumber \\
&\left| \psi _{e,m}^{\pm } \right\rangle =\left| \begin{array}{cc}
   a_{m}^{\pm }\left| m \right\rangle   \\
   b_{m+1}^{\pm }\left| m+1 \right\rangle   \\
\end{array} \right|, m\ge 0,&\nonumber \\
&\frac{{{E}_{e,0}}}{\hbar {{\omega }_{ce}}}=\frac{1}{2}-{{Z}_{e}},
\left| {{\psi }_{e,0}} \right\rangle =\left| \begin{array}{cc}
   0  \\
   \left| 0 \right\rangle   \\
\end{array} \right|, &\\
&b_{m+1}^{-}=\frac{i\alpha \sqrt{2}\sqrt{m+1}a_{m}^{-}}{\left( \frac{1}{2}-{{Z}_{e}} \right)+\sqrt{{{\left( \frac{1}{2}-{{Z}_{e}} \right)}^{2}}+2|\alpha {{|}^{2}}(m+1)}}. &\nonumber
\end{eqnarray}
In the following calculations we will take into account only two lowest states of the conduction electron $(e,{{R}_{i}})$ with $i=1,2$. One of them has the energy $E_{e}(R_{1})=\hbar {\omega }_{ce}\varepsilon _{e,0}^{-}$ and the wave function $\left| \psi _{e,0}^{-} \right\rangle $ ; the second one has the energy $E_{e}(R_{2})=\hbar {\omega }_{ce}{\varepsilon }_{e,0}$ and the wave function $\left| {{\psi }_{e,0}} \right\rangle $.
The exact solutions taking into account the nonparabolicity of the hh dispersion law have the forms [18-21]:
\begin{eqnarray}
&\frac{E_{h,m}^{\pm }}{\hbar {{\omega }_{ch}}}=\varepsilon _{h,m}^{\pm }=(m-1)+\delta ({{m}^{2}}-2m+{13}/{4}\;)\pm &\nonumber\\
&\sqrt{\left( \frac{3}{2}-{{Z}_{e}}+3\delta (m-1) \right)^{2}+8{\beta }^{2}m(m-1)(m-2)},&\nonumber \\
&m\ge 3& \nonumber \\
&\frac{{{E}_{h,m}}}{\hbar {{\omega }_{ch}}}={{\varepsilon }_{h,m}}=(m+{1}/{2}\;)+\delta {{(m+{1}/{2}\;)}^{2}}-{{Z}_{h}},& \nonumber \\
& m\le 2,&
\end{eqnarray}
with the corresponding wave functions
\begin{eqnarray}
&\left| \psi _{h,m}^{\pm } \right\rangle =\left| \begin{array}{cc}
   c_{m}^{\pm }\left| m \right\rangle   \\
   d_{m-3}^{\pm }\left| m-3 \right\rangle   \\
\end{array} \right|,m\le 3, &\nonumber \\
&\left| {\psi }_{h,m} \right\rangle =\left| \begin{array}{cc}
   \left| m \right\rangle   \\
   0  \\
\end{array} \right|,m\le 2,&
\end{eqnarray}
and the coefficient $c_{m}^{-}$ and $d_{m-3}^{-}$ satisfying the relation:
\begin{eqnarray}
&c_{m}^{-}=-2i\sqrt{2}\beta \sqrt{m(m-1)(m-2)}/ &\nonumber \\
&/[({3}/{2-{{Z}_{h}}}\;)+3\delta (m-1)+ &\nonumber \\
&\sqrt{{{\left( \frac{3}{2}-{{Z}_{e}}+3\delta (m-1) \right)}^{2}}+8{{\beta }^{2}}m(m-1)(m-2)}]& \nonumber \\
& m\ge 3.&
\end{eqnarray}
In the following calculations we will use only three hh lowest Landau states denoted as $(h,{{R}_{j}})$ with $j=1,2,3$. %First of them has the energy $E_{h}(R_{1})=\hbar {{\omega }_{ch}}\varepsilon _{h,3}^{-}$ and wave function $\left| \psi _{h,3}^{-} \right\rangle $, the second one has the energy $E_{h}(R_{2})=\hbar {{\omega }_{ch}}{{\varepsilon }_{h,0}}$ and the wave function $\left| {{\psi }_{h,0}} \right\rangle $, whereas the third state is selected with the energy $E_{h}(R_{3})=\hbar {{\omega }_{ch}}\varepsilon _{h,4}^{-}$ and with wave function $\left| \psi _{h,4}^{-} \right\rangle $.
The combination of the two electron states $(e,{{R}_{i}})$ with three hh states $(h,R_{j})$ gives rise to six states of the electron-hole pairs denoted as $F_{n}$ with $n=1,2,...6$. These combinations have the energies of the band-to-band quantum transitions $E_{cv}(F_{n})$. Being counted from the semiconductor energy gap $E_{g}^{0}$ existing in the absence of the perpendicular magnetic and electric fields their energies are
\begin{eqnarray}
&E_{cv}(F_{n})-E_{g}^{0}=E_{cv}(e,R_{i};h,R_{j})-E_{g}^{0}= &\nonumber\\
&=E_{e}(R_{i})+E_{h}(R_{j}),& \nonumber \\
&i=1,2,j=1,2,3,n=1,2,...6&
\end{eqnarray}

\section{Magnetoexcitons}
Contrary to the Wannier-Mott exciton which has an atom-like structure determined by the Coulomb electron-hole interaction, the 2D magnetoexciton has a structure similar to the electric dipole moving in the plane of the layer with the center-of-mass in-plane wave vector $\vec{k_{||}}$ , with the in-plane axis oriented perpendicular to $\vec{k_{||}}$ and with the length equal to $|\vec{k_{||}}|l_{0}^{2}$ , were $l_{0}$ is the magnetic length. Such structure is completely determined by the Lorentz force exercised by the magnetic field perpendicular to the layer. For example, if we have an e-h pair with an electron moving along the axis $x$ with wave number $p_{x}$ and a hole moving in the same direction with the wave number $q_{x}$, they will be deviated by the Lorentz force in a perpendicular in-plane direction along the axis $y$ and will effectuate quantized oscillations around the gyration points with coordinate $y_{e}=p_{x}l_{0}^2$ and $y_{h}=-q_{x}l_{0}^2$. The distance between these points $y_{e-h}=y_{e}-y_{h}=(p_{x}+q_{x})l_{0}^2=k_{x}l_{0}^2$ determine the arm of the electrical dipole moving along the $x$ axis with the center-of-mass wave vector $k_{x}$. The electron and hole wave vectors can be expressed as $p_{x}=\frac{k_{x}}{2}+t$ and $q_{x}=\frac{k_{x}}{2}-t$, where $\vec{t}$ is the wave vector of the relative motion. The Lorentz force determines not only the gyration points, which depend also on the electric charge of the particles, but also the relative e-h motion in the $y$ direction. Its wave function has the character of $\delta(y-tl_{0}^2)$ function, which in the momentum representation looks as $e^{ik_{y}tl_{0}^{2}}$. It is characterized by the second quantum number $k_{y}$, which together with $k_{x}$ form the 2D wave vector $\vec{k_{||}}(k_{x},k_{y})$, which characterizes the quantum state of the magnetoexciton.

The magnetoexcitons in the GaAs-type QWs are created by electrons in the $s$-type conduction band and by heavy holes in the $p$-type valence band, where the periodic parts of the Bloch wave functions are characterized by the magnetic quantum number $M_{h}=\pm 1$, that is equivalent to the circular polarization $\vec{\sigma}_{M_{h}}=\frac{1}{\sqrt{2}}(\vec{a}_{1}\pm i\vec{a}_{2})$, where $\vec{a}_{1}$ and $\vec{a}_{2}$ are the in-plane unit vectors. The periodic parts are not modified by the external magnetic field and coexist side by side with the envelope parts undergoing the Landau quantization described above.
The magnetoexciton wave functions are expressed through the magnetoexciton creation operator which  are expressed through the electron and hole creation operators $a_{R_{i},t}^{\dagger }$ and $b_{M_{h},R_{j},t}^{\dagger}$, as it was shown in Ref.[24-27]:
\begin{eqnarray}
&\left| {\psi }_{ex}(R_{i};M_{h},R_{j};\vec{k}_{\parallel}) \right\rangle =\hat{\psi }_{ex}^{\dagger}(\vec{k}_{\parallel};R_{i};M_{h},R_{j})\left| 0 \right\rangle , &\nonumber\\
&\hat{\psi }_{ex}^{\dagger }( \vec{k}_{\parallel};R_{i};M_{h},R_{j})=\frac{1}{\sqrt{N}}\sum\limits_{t}{e^{i{k_{y}}tl_{0}^{2}}}a_{R_{i},t+\frac{k_{x}}{2}}^{\dagger }b_{{M_{h},R_{j},-t+\frac{k_{x}}{2}}}^{\dagger}, &\nonumber\\
&a_{R_{i},t}\left| 0 \right\rangle =b_{M_{h},R_{j},t}\left| 0 \right\rangle =0, N=\frac{S}{2\pi l_{0}^{2}}&
\end{eqnarray}
Here $| 0 \rangle $ is the vacuum state and $S$ is the layer surface area. The written expressions reflect the described above picture. The creation of the magnetoexciton with the wave vector $\vec{k_{||}}\neq 0$ as a result of the optical quantum transitions requires the amount of energy smaller than the renormalized band gap $E_{cv}(F_{n})$ by the value of the ionization potential $I_{ex}(F_{n},\vec{k_{||}})$. It is caused by the Coulomb electron-hole attraction inside the magnetoexciton. The Coulomb interaction is small in comparison with the cyclotron energies and cannot change essentially the electron-exciton structure established by the Lorentz force. The magnetoexciton with zero ionization potential does not exist. The ionization potential vanishes at $|\vec{k_{||}}| \rightarrow \infty$, whereas in the range of small values of $|\vec{k_{||}}|<\frac{\pi}{L_{c}}$ its behavior determines the quadratic dispersion law of the magnetoexciton and its magnetic mass $M(F_{n},B)$. This dependence gives the value $E_{ex}(F_{n})$ with saturation-type behavior in the limit $|\vec{k_{||}}| \rightarrow \infty$.
\begin{eqnarray}
&E_{ex}(F_{n},\vec{k_{||}})=E_{cv}(F_{n})-I_{ex}(F_{n},\vec{k_{||}}), &\\
&I_{ex}(F_{n},\vec{k_{||}})=I_{ex}(F_{n},0)-\frac{\hbar^2\vec{k_{||}}^2}{2M(F_{n},B)},|\vec{k_{||}}|<\frac{\pi}{L_{c}},& \nonumber \\
&E_{ex}(F_{n},\vec{k_{||}})\approx E_{ex}(F_{n},0)+\frac{\hbar^2\vec{k_{||}}^2}{2M(F_{n},B)},& \nonumber\\
&E_{ex}(F_{n},0)=E_{cv}(F_{n})-I_{ex}(F_{n},0),& \nonumber
\end{eqnarray}
where $L_{c}$ is the cavity length. The concrete expressions for $E_{ex}$ were obtained in Ref.[28] for the all six states $F_{n}$.

The cavity photons in this description are characterized by the circular polarizations $\vec{\sigma }_{\vec{k}}^{\pm }$  oriented along the three-dimensional (3D) wave vector $\vec{k}={{\vec{a}}_{3}}{{k}_{z}}+{{\vec{k}}_{\parallel }}$  with the unit vector ${{\vec{a}}_{3}}$ perpendicular to the layer, with the quantized longitudinal component ${{k}_{z}}=\pm \frac{\pi }{{{L}_{c}}}$, where ${{L}_{c}}$ is the resonator length and with the in-plane component ${{\vec{k}}_{\parallel }}$. In a general case the cavity photons have an oblique incidence on the QW surface, that permits to excite the magnetoexcitons with ${{\vec{k}}_{\parallel }}\ne 0$. The interaction between the magnetoexciton denoted by the quantum numbers $(F_{n},\vec{k_{||}})$ and the cavity photon with circular polarization $\vec{\sigma_ {\vec{k}}}$, wave vector $\vec{k}$ with component $k_{z}$ in the frame of the GaAs band structure is expressed through the matrix elements containing all the needed information. They are:
\begin{eqnarray}
&\phi (F_{n},\vec{k}_{\parallel },M_{h};\vec{\sigma }_{\vec{k}},k_{z})=\varphi (F_{n},\vec{k}_{\parallel })\cdot (\vec{\sigma }_{\vec{k}}\cdot \vec{\sigma}_{M_{h}^{*}})&
\end{eqnarray}
where the coefficients $\varphi(F_{n},\vec{k_{||}})$ are:
\begin{eqnarray}
&\varphi(F_{1},\vec{k_{||}})=-\phi_{cv}a_{0}^{-*}d_{0}^{-*},& \nonumber \\
&|\varphi(F_{1},\vec{k_{||}})|^{2}=|\phi _{cv}|^{2}|a_{0}^{-}|^{2}|d_{0}^{-}|^{2},& \nonumber\\
&\varphi(F_{4},\vec{k_||})=\phi_{cv},| \varphi(F_{4},\vec{k_{||}})|^{2}=|\phi_{cv}|^{2}, &\nonumber\\
&\varphi(F_{3},\vec{k_{||}})=\phi_{cv}b_{1}^{-*}(\frac{-k_{x}+ik_{y}}{\sqrt{2}})l_{0},& \nonumber\\
&|\varphi(F_{3},\vec{k_{||}})|^{2}=|\phi_{cv}|^{2}|b_{1}^{-}|^{2}\frac{|\vec{k_{||}}|^{2}l_{0}^{2}}{2},& \nonumber\\
&\varphi(F_{5},\vec{k_{||}})=-\phi_{cv}a_{0}^{-*}d{0}^{-*}(\frac{k_{x}+ik_{y}}{\sqrt{2}})l_{0},& \nonumber\\
&|\varphi(F_{5},\vec{k_{||}})|^{2}=|\phi_{cv}|^{2}|a_{0}^{-}|^{2}| d_{0}^{-}|^{2}\cdot \frac{|\vec{k_{||}}|^{2}l_{0}^{2}}{2}, &\nonumber\\
&\phi_{cv}=\frac{e}{m_{0}l_{0}}\sqrt{\frac{\hbar{n_{c}}}{\pi c}}{P_{cv}}(0),|\phi _{cv}|^{2}=(\frac{\hbar{n_{c}}}{\pi c})| \frac{eP_{cv}(0)}{m_{0}l_{0}}|^{2}, &\nonumber\\
&f_{osc}=|\frac{\phi_{cv}}{\hbar{\omega_{c}}}|^{2}=\frac{\hbar{n_{c}}}{\pi c}\cdot | \frac{eP_{cv}(0)}{m_{0}l_{0}\hbar{\omega_{c}}}|^{2}&
\end{eqnarray}
As one can see this interaction depends on many factors. One of them is the matrix element $P_{cv}(0)$ which determines the inter-band optical quantum transition in the GaAs-type crystal. It depends on the periodic parts of the electron Bloch wave functions which remain unchanged in the presence of the external magnetic field. It is an intrinsic selection rule related with the crystal band structure. Another selection rule arises from the Landau quantization wave functions. It leads to existence of two dipole-active quantum transitions in the states $F_{1}$ and $F_{4}$ where the quantum numbers of the Landau quantization of electron $n_{e}$ and of hole $n_{h}$ coincide $n_{e}=n_{h}$. The same selection rule leads to the two quadrupole-active quantum transitions in the state $F_{3}$ and $F_{5}$, where the quantum numbers of the Landau quantization of electron and hole differ by one and satisfy to the condition $n_{e}=n_{h}\pm 1$. Looking at the expressions (11) one can observe that the coefficient $\varphi(F_{1},\vec{k_{||}})$ and $\varphi(F_{4},\vec{k_{||}})$ describing the dipole-active transitions do not vanish in the point $\vec{k_{||}}=0$. On the contrary, the coefficients $\varphi(F_{3},\vec{k_{||}})$ and $\varphi(F_{5},\vec{k_{||}})$ corresponding to quadrupole transitions depend essentially on the components $(\pm k_{x}+ik_{y})$ and vanish in the point $\vec{k_{||}}=0$. There are another two forbidden magnetoexciton states with their coefficients $\varphi(F_{2},\vec{k_{||}})$ and $\varphi(F_{6},\vec{k_{||}})$ equal to zero. Another factors of the expressions (10), (11) have the geometric origin and are expressed through the scalar products $(\vec{\sigma_{k}^{\pm}}\cdot \vec{\sigma_{M_{h}}^{*}})$ between two vectors of circular polarizations. One of them describes the photon and another one the magnetoexciton states. The scalar products and their squared modulus depend essentially on the values of the photon wave vectors:
\begin{eqnarray}
&\vec{k_{\uparrow, \downarrow}}=\pm \vec{a_{3}}\frac{\pi}{L_{c}}+\vec{k_{||}}&
\end{eqnarray}

\section{Dispersion equation and magnetoexciton-polariton branches}
We consider the particular case of the cavity photons with the circular polarizations $\sigma _{\vec{k}_{\uparrow }}^{-}$ or $\sigma _{\vec{k}_{\downarrow }}^{+}$ and the cavity mode energy $\hbar {\omega_ {c}}$ tuned to the magnetoexciton energy level $E_{ex}(F_{1},B,0)$ , introducing the following denotations
\begin{eqnarray}
&\hbar \omega =\hbar {\omega_{c}}+E,\frac{\hbar \omega }{\hbar {\omega_{c}}}=1+\varepsilon ,\varepsilon = \frac{E}{\hbar {\omega_{c}}},& \nonumber\\
&\hbar {\omega_{\vec{k}}}=\hbar {\omega_{c}}\left(1+\frac{x^{2}}{2}\right),x=\frac{|\vec{k}_{||}|L_{c}}{\pi},& \nonumber\\
&\hbar {\omega_{c}}=\frac{E_{ex}(F_{1},B,0)}{1-{\delta }_{1}},{{\delta }_{1}}=-0.01&
\end{eqnarray}
Then the selected dispersion equation takes the form
\begin{eqnarray}
\left( \varepsilon -\frac{{{x}^{2}}}{2} \right)={{f}_{osc}}\cdot \left\{ \frac{{{\left| a_{0}^{-} \right|}^{2}}{{\left| d_{0}^{-} \right|}^{2}}\left( 1-\frac{{{x}^{2}}}{2}+\frac{7}{16}{{x}^{4}} \right)}{\left( \varepsilon +{{\delta }_{1}}-\frac{n_{c}^{2}\hbar {{\omega }_{c}}}{M\left( {{F}_{1}},B \right){{c}^{2}}}\cdot \frac{{{x}^{2}}}{2} \right)}+ \right. \nonumber \\
+\frac{\frac{{{x}^{4}}}{16}}{\left( \varepsilon +1-\frac{{{E}_{ex}}\left( {{F}_{4}},B,0 \right)}{{{E}_{ex}}\left( {{F}_{1}},B,0 \right)}\left( 1-{{\delta }_{1}} \right)-\frac{n_{c}^{2}\hbar {{\omega }_{c}}}{M\left( {{F}_{4}},B \right){{c}^{2}}}\cdot \frac{{{x}^{2}}}{2} \right)}+ \nonumber\\
+\frac{{{\left| b_{1}^{-} \right|}^{2}}{{\left( \frac{\pi {{l}_{0}}}{{{L}_{c}}} \right)}^{2}}\frac{{{x}^{6}}}{32}}{\left( \varepsilon +1-\frac{{{E}_{ex}}\left( {{F}_{3}},B,0 \right)}{{{E}_{ex}}\left( {{F}_{1}},B,0 \right)}\left( 1-{{\delta }_{1}} \right)-\frac{n_{c}^{2}\hbar {{\omega }_{c}}}{M\left( {{F}_{3}},B \right){{c}^{2}}}\cdot \frac{{{x}^{2}}}{2} \right)}+ \nonumber\\
\left. +\frac{{{\left| a_{0}^{-} \right|}^{2}}{{\left| d_{0}^{-} \right|}^{2}}{{\left( \frac{\pi {{l}_{0}}}{{{L}_{c}}} \right)}^{2}}\frac{{{x}^{2}}}{2}\left( 1-\frac{{{x}^{2}}}{2}+\frac{7}{16}{{x}^{4}} \right)}{\left( \varepsilon +1-\frac{{{E}_{ex}}\left( {{F}_{5}},B \right)}{{{E}_{ex}}\left( {{F}_{1}},B \right)}\left( 1-{{\delta }_{1}} \right)-\frac{n_{c}^{2}\hbar {{\omega }_{c}}}{M\left( {{F}_{5}},B \right){{c}^{2}}}\cdot \frac{{{x}^{2}}}{2} \right)} \right\}
\end{eqnarray}
Figure 1 shows the results of numerical solution of the equation (14).
From  Fig. 1 one can observe that the bare cavity photon branch is intersected by the branches of the bare magnetoexciton states $F_{n}$. Only the interaction of the magnetoexciton branch $F_{1}$ with the cavity photons of the selected circular polarization $\vec{\sigma}_{\vec{k} \uparrow }^{-}$ or $\vec{\sigma}_{\vec{k} \downarrow}^{+}$ gives rise to the well distinguishable polariton pictures. For another intersection points denoted as $(n-c)$ with $n=3,4,5$ the magnetoexciton-photon interactions are very small. The reconstructed polariton curves in the vicinities of these points are shown only qualitatively. The real values of the splittings arising due to the anticrossings in these points are denoted as $\Delta(n-c)$. The coordinates of these intersection points $X(n-c)$ and $Y(n-c)$ together with the corresponding splittings are explained in the figure caption. As was mentioned above they correspond to the cavity photon circular polarization $\vec{\sigma }_{\vec{k}_{\uparrow }}^{-}$ or $\vec{\sigma }_{k_{\downarrow }}^{+}$. Changing only the sign of the $k_{z}$ component of the vectors $\vec{k}_{\uparrow}$ and $\vec{k}_{\downarrow}$ we will obtain the polariton branches picture in the case of the photon circular polarizations $\vec{\sigma}_{\vec{k}_{\downarrow }}^{-}$ or $\vec{\sigma }_{\vec{k}_{\uparrow }}^{+}$.

The new energy spectrum obtained at these circular polarizations has another splittings $\Delta (n-c)$ in the intersection points with the coordinates $X(n-c)$ and Y(n-c): (a) $B=20$T, $g_{h}=5$; $X(1-c)=0.142007$, $Y(1-c) = 1.0100052$, $\Delta=9.06823\cdot10^{-6}$,
$X(3-c)=0.228535$, $Y(3-c)=1.0297925$, $\Delta=11.2865\cdot10^{-6}$, $X(4-c)=0.263951$, $Y(4-c)=1.0349685$, $\Delta=82.7071\cdot10^{-6}$. (b) $B=40$T, $g_{h}=5$; $X(1-c)=0.14038$, $Y(1-c)=1.00986762$, $\Delta=13.6447\cdot10^{-6}$, $X(3-c)=0.339592$, $Y(3-c)=1.0603962$, $\Delta=15.9892\cdot10^{-6}$, $X(4-c)=0.240025$, $Y(4-c)=1.0290947$, $\Delta=114.079\cdot10^{-6}$. (c) $B=20$T, $g_{h}=-5$; $X(1-c)=0.140703$, $Y(1-c)=1.0100376$, $\Delta=12.0272\cdot10^{-6}$,
$X(3-c)=0.25311$, $Y(3-c)=1.0355721$ $\Delta=14.71\cdot10^{-6}$, $X(4-c)=0.345856$, $Y(4-c)=1.0592373$, $\Delta=109.109\cdot10^{-6}$. (d) $B=40$T, $g_{h}=-5$; $X(1-c)=0.141076$, $Y(1-c)=1.00989149$ $\Delta=13.6789\cdot10^{-6}$, $X(3-c)=0.37412$, $Y(3-c)=1.0729855$, $\Delta=15.9056\cdot10^{-6}$, $X(4-c)=0.37892$, $Y(4-c)=1.0716257$, $\Delta=111.541\cdot10^{-6}$.
The considerable changes in comparison with the previous case appeared at the points $(1-c)$ and $(4-c)$, that demonstrate the presence of the significant gyrotropy effect. The Zeeman splitting effects can be observed by changing the value of the hh $g-$ factor from the value 5 to -5. The changes of the polariton energy spectrum do not exceed 18mev in the case of $B=40$T and 8.5 mev at $B=20$T.
\begin{figure}[h]
\includegraphics[scale=0.045]{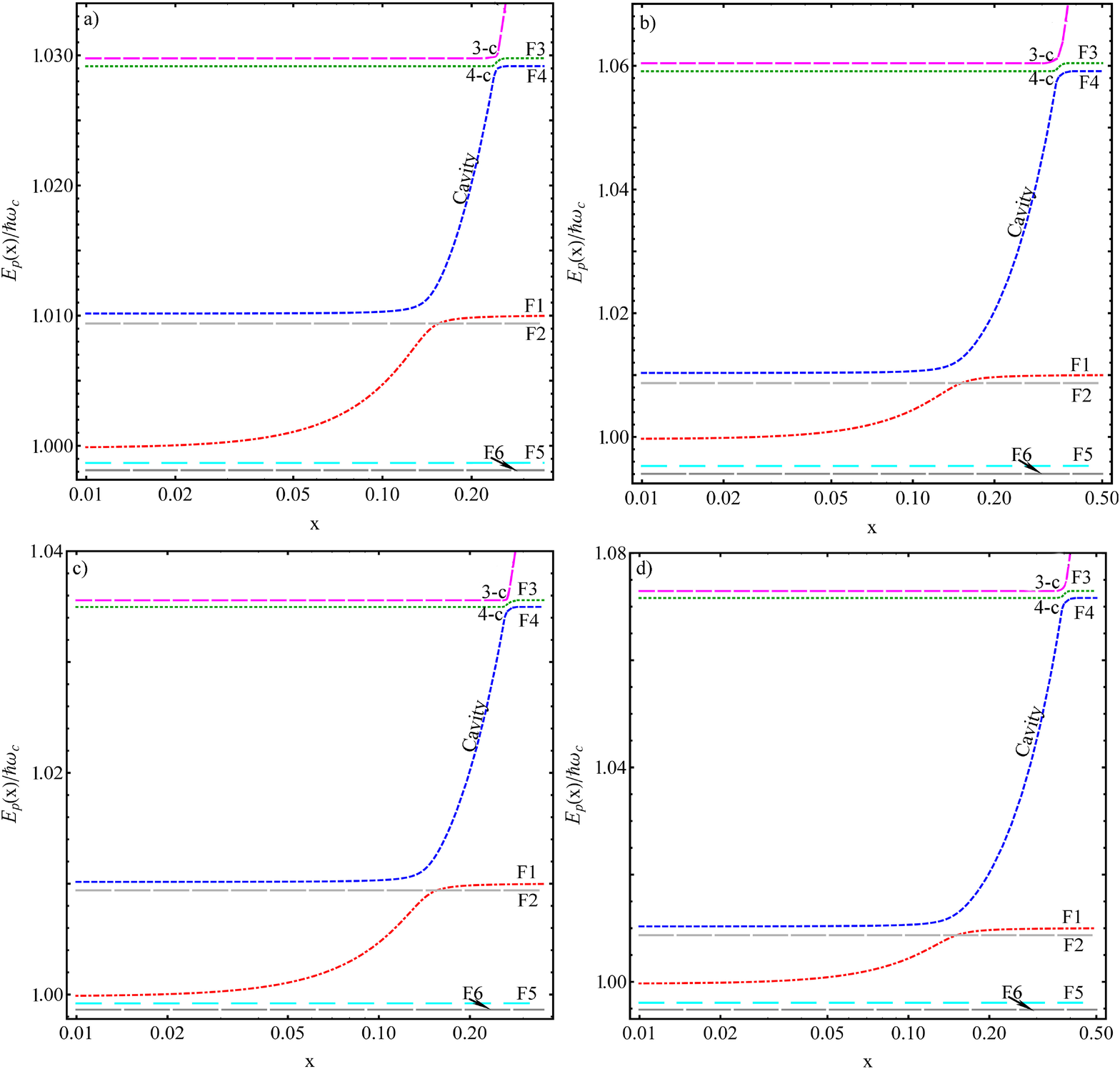}
\caption{Five cavity magnetoexciton-polariton energy branches in dependence on the dimensionless in-plane wave number $x$ arising as result of the interaction of the cavity photons with circular polarizations $\vec{\sigma }_{\vec{k}_{\uparrow }}^{-}$ or $\vec{\sigma }_{\vec{k}_{\downarrow }}^{+}$ with four magnetoexciton energy branches $F_{1}$, $F_{4}$, $F_{3}$ and $F_{5}$ characterized by the circular polarizations $\vec{\sigma}_{\pm 1}$ , two branches being dipole-active and two quadrupole active, when the cavity mode energy is tuned to the magnetoexciton energy level $E_{ex}(F_{1},B,0)$. The influence of the external electric field and of the RSOC with the parameters $E_{z}=30$kV/cm and $C=20$ is represented in four variants. They appear from the combinations of two values of the magnetic field strength $B$ and of two values of the hh $g-$factor  $g_{h}$ as follows:  (a) $B=20$T, $g_{h}=5;$ (b) $B=40$T, $g_{h}=5;$ (c) $B=20$T, $g_{h}=-5;$ (d) $B=40$T, $g_{h}=-5$. The electron $g-$factor was selected $g_{e}=1$. For the completeness two forbidden magnetoexciton energy branches $F_{2}$ and $F_{6}$ are added. The intersection points and the splittings of the polariton curves in these points are: (a)$X(1-c)=0.145$, $Y(1-c)=1.01015$, $\Delta=180.17\cdot10^{-6}$; $X(4-c)=0.241$, $Y(4-c)=1.02919$, $\Delta=11.23\cdot 10^{-6}$; $X(3-c)=0.245$, $Y(3-c)=1.02980$, $\Delta=11.29\cdot 10^{-6}$;
(b)$X(1-c)=0.146$, $Y(1-c)=1.01019$, $\Delta=359.66\cdot10^{-6}$; $X(4-c)=0.343$, $Y(4-c)=1.05920$, $\Delta=15.74\cdot 10^{-6}$; $X(3-c)=0.347$, $Y(3-c)=1.06040$, $\Delta=16\cdot10^{-6}$;
(c)$X(1-c)=0.143$, $Y(1-c)=1.01005$, $\Delta=183.96\cdot10^{-6}$; $X(4-c)=0.264$, $Y(4-c)=1.03492$, $\Delta=11.19\cdot10^{-6}$; $X(3-c)=0.268$, $Y(3-c)=1.03561$, $\Delta=11.26 \cdot 10^{-6}$;
(d)$X(1-c)=0.145$, $Y(1-c)=1.01013$, $\Delta=366.13\cdot10^{-6}$; $X(4-c)=0.377$, $Y(4-c)=1.07161$, $\Delta=15.60 \cdot 10^{-6}$; $X(3-c)=0.380$, $Y(3-c)=1.07287$, $\Delta=15.91\cdot10^{-6}$}
\end{figure}

The developed theoretical model, which takes into account the electron structure of the 2D magnetoexcitons due to the influence of the  strong perpendicular magnetic and electric fields, is applicable for the GaAs-type QWs only in the range of magnetic field strength $B \ge 10$T. A similar variant but without RSOC was considered in Refs.[15, 16]. The experimental investigations of the cavity exciton-polaritons in a range of small and strong magnetic fields up to 14T were reported in  Ref.[29]. The theoretical model for this case was developed in Ref.[30]. A short review of the comprehensive experimental findings is given in Ref.[29] and will be compared with the theoretical results obtained in the present study. The experiments in Ref.[29] were performed using the 8 nm-thick In$_{0.04}$Ga$_{0.96}$As QW embedded into the resonator with the distributed Bragg reflectors. The Wannier-Mott excitons have a resonance energy 1.484 ev,  the exciton binding energy equal approximately 7 mev, and the effective mass $0.046m_{0}$. The polariton linewidth was equal to 0.3mev. These objects are interesting due to the BEC phenomenon with the participation of the cavity polaritons accompanied by the formation of superfluidity with zero viscosity was observed, and due to the development of the polariton-based devices [31]. There are different effects of the magnetic field on the polaritons revealed in Ref.[29] already in the linear regime. The magnetic field induces the changes in the linewidth. The interpretation of this result on the base of the hydrogen interpolation model [30] takes into account the exciton energy shift including the Zeeman splitting, the modification of the exciton-photon coupling strength and the change of the Rabi splitting. The magnetic field is considered as a parameter changing the tuning of the exciton-photon resonances. The changes of the emission intensity and of the polariton linewidth were explained in Ref.[29] by the magnetically modified scattering processes with participation of the acoustic phonons. The increase of the exciton oscillator strength $f_{osc}$ was directly determined in Ref.[29] measuring the Rabi splitting $\Omega $ of a polariton branches. Taking into account the relation $\Omega \sim \sqrt{{Nf_{osc}}/{L_{eff}}\;}$, where $N$ is the number of QWs embedded into the resonator and $L_{eff}$ is the effective length, the $f_{osc}$ was calculated. These results are depicted  in the fig. 6 of Ref.[29]. One can see that the  value of $\hbar \Omega $ increases from 3.4 mev to 4.8 mev at $B=14$ T, whereas the $f_{osc}$ increases by a factor of two.

In the present model the dependence of the Rabi frequency $\Omega $ on the magnetic field strength is determined by the relation $\Omega \sim l_{0}^{-1}\sim \sqrt{B}$, where $l_{0}$ is the magnetic length. It leads to the dependence for the oscillator strength $f_{osc}\sim |\Omega |^{2}\sim B$.
Another property of the cavity polariton spectra is the Zeeman splitting of the lower polariton branch. Two components of the bright exciton level are decoupled and couple independently with the photon of corresponding circular polarization. The observed experimentally in Ref.[29] the Zeeman splitting of the lower polariton branch reveals the decrease in the dependence on the magnetic field strength up to 14T, instead of the expected increase.
In the frame of our model the magnetoexciton states F1-F6 are represented by the non-degenerate branches with the evantail-type dependence on the magnetic field strength with the increasing splitting between them in the absence of the RSOC.
Only in the presence of the RSOC, as well as taking into account the nonparabolicity of the heavy-hole dispersion law, the intersections and overlapping between the magnetoexciton branches appear. The surprising behavior of the Zeeman splitting revealed in  Ref.[29] is possibly  related with the dependence of the electron and hole g-factors $g_{e}(B)$ and $g_{h}(B)$ on the magnetic field strength B [32-34].

The exciton-polaritons considered above are due to the inter-band optical quantum transitions and have the exciton frequency within  the optical range of spectrum. It is much greater than the Rabi frequency determining the exciton-photon interaction constant. In contrast to this, if we  deal with the intra-band [35, 36] or with the intersubband [37] quantum transitions the situation can be changed considerably so that the Rabi frequency $\Omega_{R}$ can be of the same order or even greater than the excitation frequency $\omega_{0}$. Such special conditions were investigated in Ref.[35-37]. This case is named as the ultrastrong coupling regime. As it was mentioned in Ref.[36] it permits to manipulate the cavity ground state, to modify the decoherence properties and to enhance the nonadiabatic cavity QED effects.  The two-dimensional electron gas in the semiconductor conduction band with the massive electrons and the quadratic dispersion law subjected to the action of the perpendicular magnetic field was considered in Ref.[35]. The Landau quantization levels were supposed to be filled with a great integer filling factor $\nu$. The excitation quantum transitions take place between the last occupied Landau level with $n=\nu$ and the first unoccupied level $n=\nu +1$ and the excitation frequency coincides with the cyclotron frequency. It was shown that the ratio of the Rabi frequency to the excitation frequency equals to $\frac{\Omega_{R}}{\omega_{0}}=\sqrt{\alpha n_{QW} \nu}$, where $\alpha$ is the fine structure constant $\alpha=\frac{e^{2}}{\hbar c}=\frac{1}{137}$ and $n_{QW}$ is the number of the QWs embedded into the resonator.  The desirable ultrastrong coupling regime was achieved for the  special conditions with $n_{QW}\sim\nu\sim100$, that was confirmed qualitatively by the experimental result in the THz domain [38]. The same regime was investigated in Ref.[36] in the case of graphen, so as to clarify the influence of the Dirac cone-type dispersion law of the massless electron when the Landau quantization levels have the dependence $E_{n}=\hbar \omega_{0}\sqrt{n}$ with $\omega_{0}=\frac{V_{F}\sqrt{2}}{l_{0}}$, where $V_{F}$ is the Fermi velocity. The case of the quantum transitions between the conduction subbands of the asymmetric QW was considered in  Ref[37], taking into account the Rashba [1] as well as the Dresselhaus [39] spin-orbit coupling. The spin-conserving intersubband quantum transitions are characterized by the selection rules allowing the absorption only of the transverse magnetic polarized (TM) electromagnetic mode.

The electromagnetic wave moving inside the 3D space of the microcavity with embedded in it QW has a wave vector $\vec{k}=\vec{a}_{3}k_{z}+\vec{k}_{||}$ with in-plane component $\vec{k}_{||}=\vec{a}_{1}k_{x}+\vec{a}_{2}k_{y}$. The two vectors of linear polarizations $\vec{S}_{\vec{k}}$ and $\vec{t}_{\vec{k}}$ as well as of the circular polarizations $\vec{\sigma}_{\vec{k}}^{\pm}$ chosen in the form satisfying the orthogonality and normalization conditions are:
\begin{eqnarray}
&\vec{S}_{\vec{k}}=\vec{a}_{3}\frac{|\vec{k}_{||}|}{|\vec{k}|}-\frac{\vec{k}_{||}}{|\vec{k}_{||}|}\cdot\frac{k_{z}}{|\vec{k}|}=\vec{S}_{\vec{-k}}, \vec{t}_{\vec{k}}=\frac{\vec{a}_{1}k_{x}-\vec{a}_{2}k_{y}}{|\vec{k}_{||}|}=-\vec{t}_{\vec{-k}},& \nonumber \\
&(\vec{S}_{\vec{k}}\cdot\vec{k})=(\vec{t}_{\vec{k}}\cdot \vec{k})=(\vec{S}_{\vec{k}}\cdot\vec{t}_{\vec{k}})=0,& \nonumber \\ &\vec{\sigma}_{\vec{k}}^{\pm}=\frac{1}{\sqrt{2}}(\vec{S}_{\vec{k}}\pm i\vec{t}_{\vec{k}}), (\vec{\sigma}_{\vec{k}}^{\pm}\cdot\vec{\sigma}_{\vec{k}}^{\pm*})=1,& \nonumber \\
&(\vec{\sigma}_{\vec{k}}^{\pm}\cdot\vec{\sigma}_{\vec{k}}^{\mp*})=0,|\vec{S}_{\vec{k}}|=|\vec{t}_{\vec{k}}|=|\vec{\sigma}_{\vec{k}}^{\pm}|=1.&
\end{eqnarray}
The circular polarizations are preferable quantum numbers in the presence of a static perpendicular magnetic field $\vec{B}||\vec{a}_{3}$ when the light wave vector $\vec{k}$ has a comparatively small in-plane component $|\vec{k}_{||}|<\frac{\pi}{L_{c}}$ as it was considered in our paper. In a general case the linear polarization  is more suitable. In this case the electromagnetic wave with electric and magnetic vectors $\vec{E}_{\vec{k}}$ and $\vec{H}_{\vec{k}}$ oriented along the polarization vectors $\vec{t}_{\vec{k}}$ and $\vec{S}_{\vec{k}}$  correspondingly has the properties of the surface TE wave for the particles moving in-plane of the QW with the wave vectors $\vec{k}_{||}$. Indeed in this case:
\begin{eqnarray}
&\vec{E}_{\vec{k}}||\vec{t}_{\vec{k}}\perp \vec{k}_{||}, \vec{H}_{\vec{k}}|| \vec{S}_{\vec{k}}=\vec{H}_{\vec{k}T}+\vec{H}_{\vec{k}L},& \\
&\vec{H}_{\vec{k}T}||\vec{a}_{3}\perp \vec{k}_{||}, \vec{H}_{\vec{k}L}||\vec{k}_{||}, \vec{E}_{\vec{k}}+\vec{E}_{-\vec{k}}=0, \vec{H}_{\vec{k}}=\vec{H}_{-\vec{k}}&.  \nonumber
\end{eqnarray}
In the opposite case the 3D linear polarized wave has the properties of the surface TM wave, where:
\begin{eqnarray}
&\vec{E}_{\vec{k}}||\vec{S}_{\vec{k}}=\vec{E}_{\vec{k}T}+\vec{E}_{\vec{k}L},\vec{E}_{\vec{k}T}||\vec{a}_{3}\perp \vec{k}_{||},&  \\
&\vec{E}_{\vec{k}L}||\vec{k}_{||},\vec{E}_{\vec{k}}=\vec{E}_{-\vec{k}}, \vec{H}_{\vec{k}}|| \vec{t}_{\vec{k}}\perp \vec{k}_{||}, \vec{H}_{\vec{k}}+\vec{H}_{-\vec{k}}=0. & \nonumber
\end{eqnarray}
In such a way the 3D linearly polarized light wave acts on the 2D magnetoexcitons similarly to a surface TE or TM waves.
\section{Conclusions}
The properties of the 2D cavity polaritons under the influence of the strong perpendicular magnetic and electric fields giving rise to the LQ of the 2D electron and hole gases accompanied by the RSOC and by the ZS were investigated. The obtained exact solutions were used to determine the creation energies of the 2D magnetoexcitons as well as of their interaction with cavity photons. The fifth order dispersion equation was solved taking into account two dipole-active and two-quadrupole active magnetoexciton branches interacting with the cavity photon branch. The Rabi splitting $\Omega_{R}$ of the magnetoexciton-polariton branches and the magnetoexciton oscillator strength depend on the magnetic field strength as $\sqrt{B}$ and $B$ correspondingly.
We can conclude that the experimental and theoretical investigations of the 2D cavity polaritons in the range of high magnetic fields are in progress.

I.V.P. and E.V.D. thanks the Foundation for Young Scientists of the Academy of Sciences of Moldova for financial support (14.819.02.18F).

\end{document}